\documentclass[a4paper,aps,11pt,showpacs,preprint]{revtex4}
\usepackage{graphicx,times,nicefrac,amsmath,amsfonts,amssymb,amsthm,epsf,bm,bbm,longtable,dcolumn}
\usepackage{revsymb,wasysym,mathrsfs,color}
\DeclareGraphicsExtensions{.pdf,.jpg}

\begin{document}

\title{Main Magnetic Focus Ion Source: II. The first investigations at 10 keV}

\author{V. P. Ovsyannikov}\thanks{\url{http://mamfis.net/ovsyannikov.html}}
\affiliation{Hochschulstr. 13, D-01069  Dresden, Germany}

\author{A. V. Nefiodov}\thanks{Electronic mail: anef@thd.pnpi.spb.ru}
\affiliation{Petersburg Nuclear Physics Institute, 188300 Gatchina, St.~Petersburg, Russia}

\date{Received \today}
\widetext
\begin{abstract}
The basic principles of design for the compact ion source of new generation are presented. The device uses the local ion trap created by the axial electron beam rippled in a thick magnetic lens. In accordance with this feature, the ion source is given the name \emph{main magnetic focus ion source}. The experimental evidences for the production of Ir$^{59+}$, Xe$^{44+}$, and Ar$^{16+}$ ions are obtained. The control over depth of the local ion trap is shown to be feasible.
\end{abstract}
\maketitle

\section{Introduction}

The electron current density is the key physical parameter for the production of highly charged ions in the electron beam. In the standard electron beam ion sources and traps (EBIS/Ts), the electrostatic ion trap is used. This trap is created by the electron-optical system, which consists of the electron gun, drift tubes with positive potential barriers at the edge sections, and the electron collector. The length of electron beam is large enough. The ultimate current density in the long and relatively smooth electron beam is limited by the Brillouin conditions.

By contrast, the electron current density in the local ion trap created by rippled electron beam can be much higher \cite{1,2}. The length of the local trap is very small. Accordingly, the device, which operates by using this method of ionization, can be small in comparison with the standard EBIS/Ts. The novel ion source is labelled as main magnetic focus ion source (MaMFIS). In this paper, we present the general design for the pilot example of MaMFIS as well as the experimental results obtained at the electron beam energy of 10 keV. The objective of the experimental study is the test of theoretical predictions for ionization processes in the local ion trap \cite{2}.

\section{Formation of local ion trap}

The high-charge states of ions confined in the local ion trap are produced due to multiple sequential ionization by the electron impact. The local ion trap is located in crossover of the rippled electron beam propagating through the cylindrical drift tube. The electron beam is rippled by the thick magnetic lens, which focuses the electron beam projected from the spot-sized cathode into a sequence of the crossovers (sharp focuses). Since the most acute magnetic focus characterized by the highest electron current density is called main, it is employed in the name for the ion source.

\begin{figure}[tp]
\includegraphics[width=0.85\columnwidth]{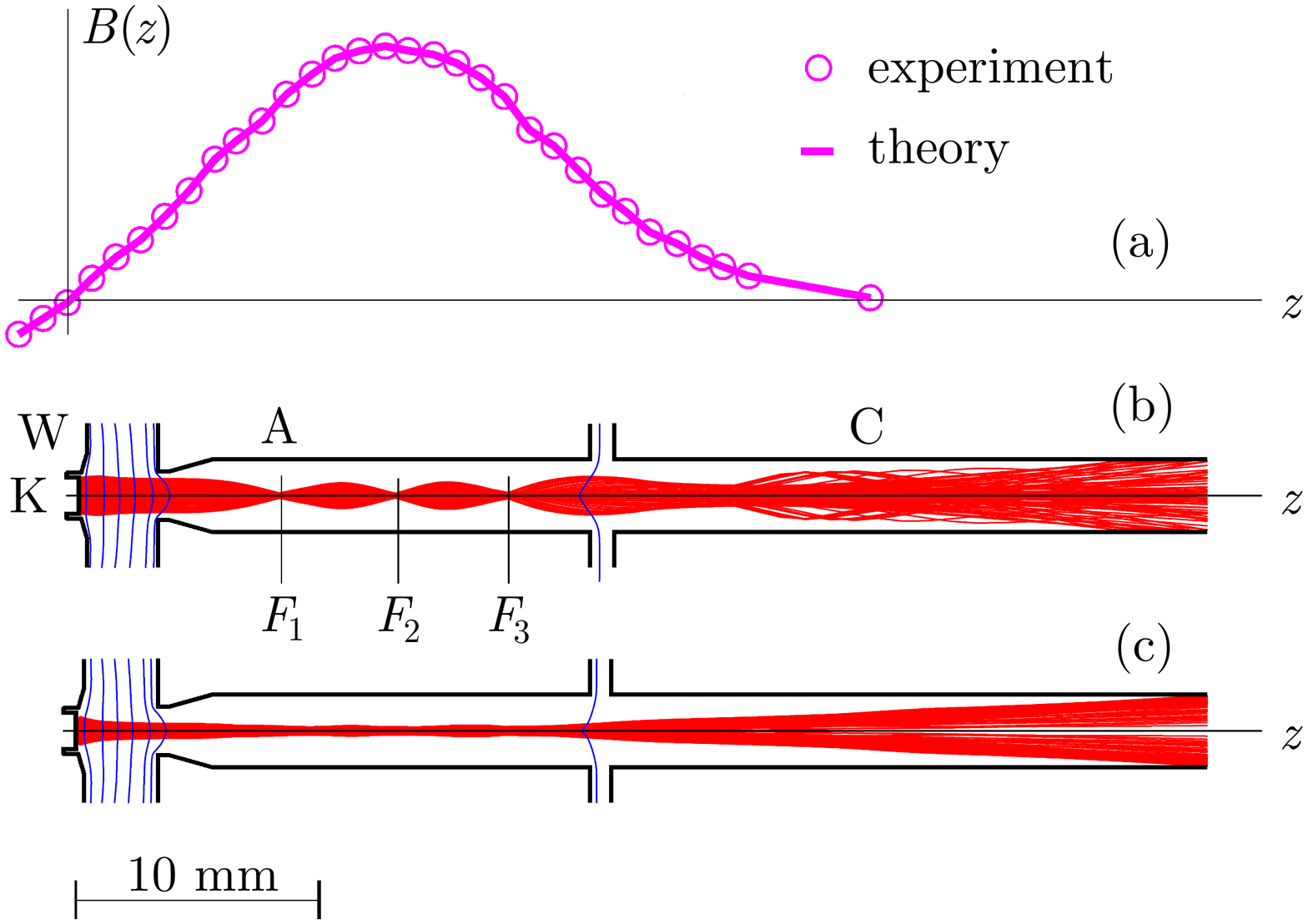}
\caption{\label{fig1} (Color online) MaMFIS-10: axial distribution of the focusing magnetic field (a),
schemes of the electron optics and electron trajectories for the trapping mode (b) and extraction mode (c).
The ion source consists of the cathode (K), Wehnelt's electrode (W), anode (A), and collector (C). The first, second, and third focuses are denoted as $F_1$, $F_2$, and $F_3$, respectively.}
\end{figure}

For axially symmetric electron beam with the radius varying from the maximum $r_\mathrm{max}$ to the minimum $r_\mathrm{min}$, the depth of the local potential well $\Delta U_\mathrm{trap}$ is given by \cite{2}
\begin{equation}
\Delta U_\mathrm{trap}= \frac{U P}{2 \pi \varepsilon_0 \sqrt{2 \eta}} \ln \frac{r_\mathrm{max}}{r_\mathrm{min}}  . \label{eq1}
\end{equation}
Here $\varepsilon_0$ is the permittivity of free space, $\eta = e/m$ is the magnitude of the electron charge-to-mass ratio, $U$ is the potential of the drift tube relative to the potential of the cathode, $P= I_{e}/U^{3/2}$ is the perveance of the electron beam, and $I_{e}$ is the electron current. In the case of the smooth electron beam ($r_\mathrm{max}=r_\mathrm{min}$), the local ion trap does not appear ($\Delta U_\mathrm{trap} = 0$).

For pilot example of the MaMFIS, the electron-optical system for the electron beam characterized by the current $I_e$ of 50 mA and the energy $E_e=eU$ of 10 keV is chosen. The magnetic field $B(z)$ is distributed over the length of about 30 mm. The axial distribution of the focusing magnetic field provides transformation of the electron beam into a sequence of three sharp focuses (see Fig.~\ref{fig1}). The magnetic field strength reaches 4.2 kG at the maximum.

\begin{figure}[tp]
\centering
\includegraphics[width=0.7\columnwidth]{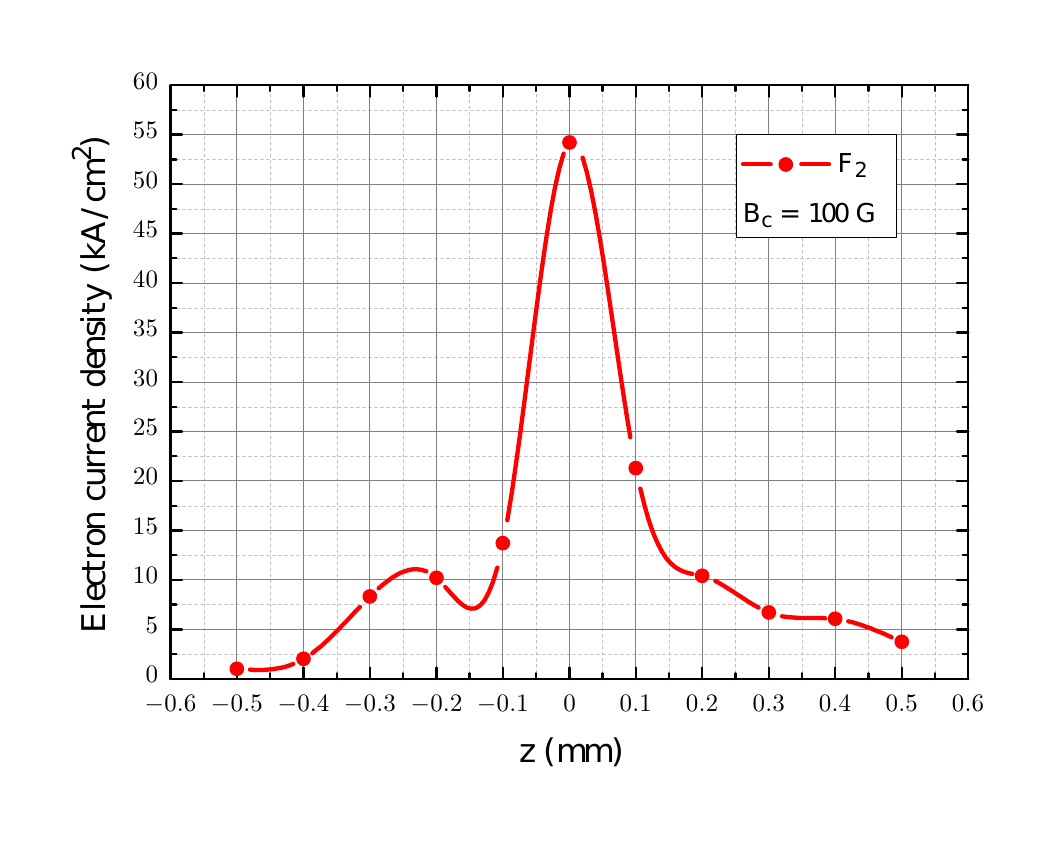}
\caption{\label{fig2} (Color online) Axial distribution of the electron current density near the second focus. The point $z =0$ corresponds to the position of the highest current density.}
\end{figure}

\begin{figure}[tp]
\centering
\includegraphics[width=0.7\columnwidth]{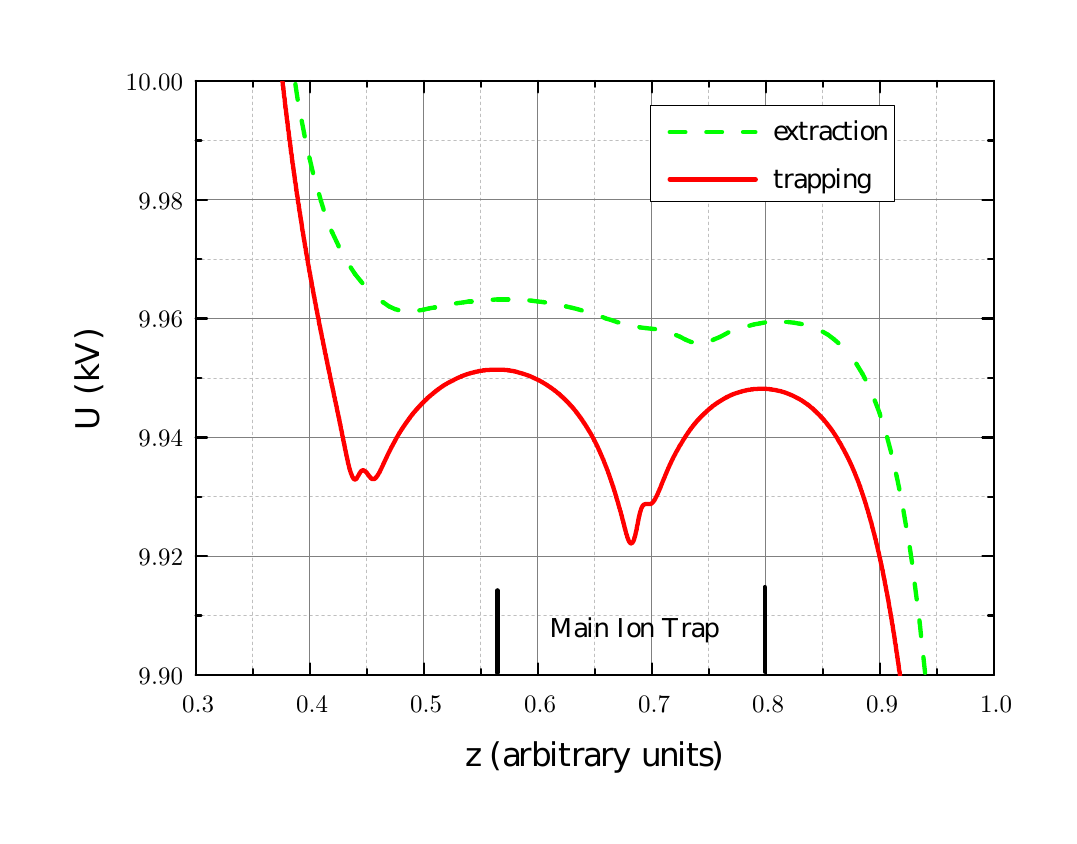}
\caption{\label{fig3} (Color online) Axial potential distribution. In extraction mode, the potential of the focusing electrode is less by 150 V than that of the cathode.}
\end{figure}

The second focus $F_2$ of the electron beam is designed for the main ion trap. The trap is located in the middle plane of installation between the magnetic coils. The distribution of the electron current density $j_e(z)$ within 1 mm around the second focus is shown in Fig.~\ref{fig2}. The amplitude value of the current density is about 55  kA/cm$^2$. The current density averaged over the length of 1 mm can be estimated to be of about 10 kA/cm$^2$.

The potential distribution along the $z$ axis of the ion source is shown in Fig.~\ref{fig3} for both trapping and extraction running modes. The rippled electron beam creates two potential wells (see the solid curve). The second well is used as the main ion trap. The depth of the trap is about 30 V. In this case, the potential of the Wehnelt's electrode is equal to the cathode potential. When potential of the focusing electrode becomes less than the potential of the cathode, the potential distribution is transformed into relatively smooth function (see the dashed curve), so that the ions can escape from the trap. The time of ion confinement is limited by the ionization time in the trapping mode.

\section{Design of MaMFIS-10}

\begin{figure}[tp]
\includegraphics[width=0.45\columnwidth]{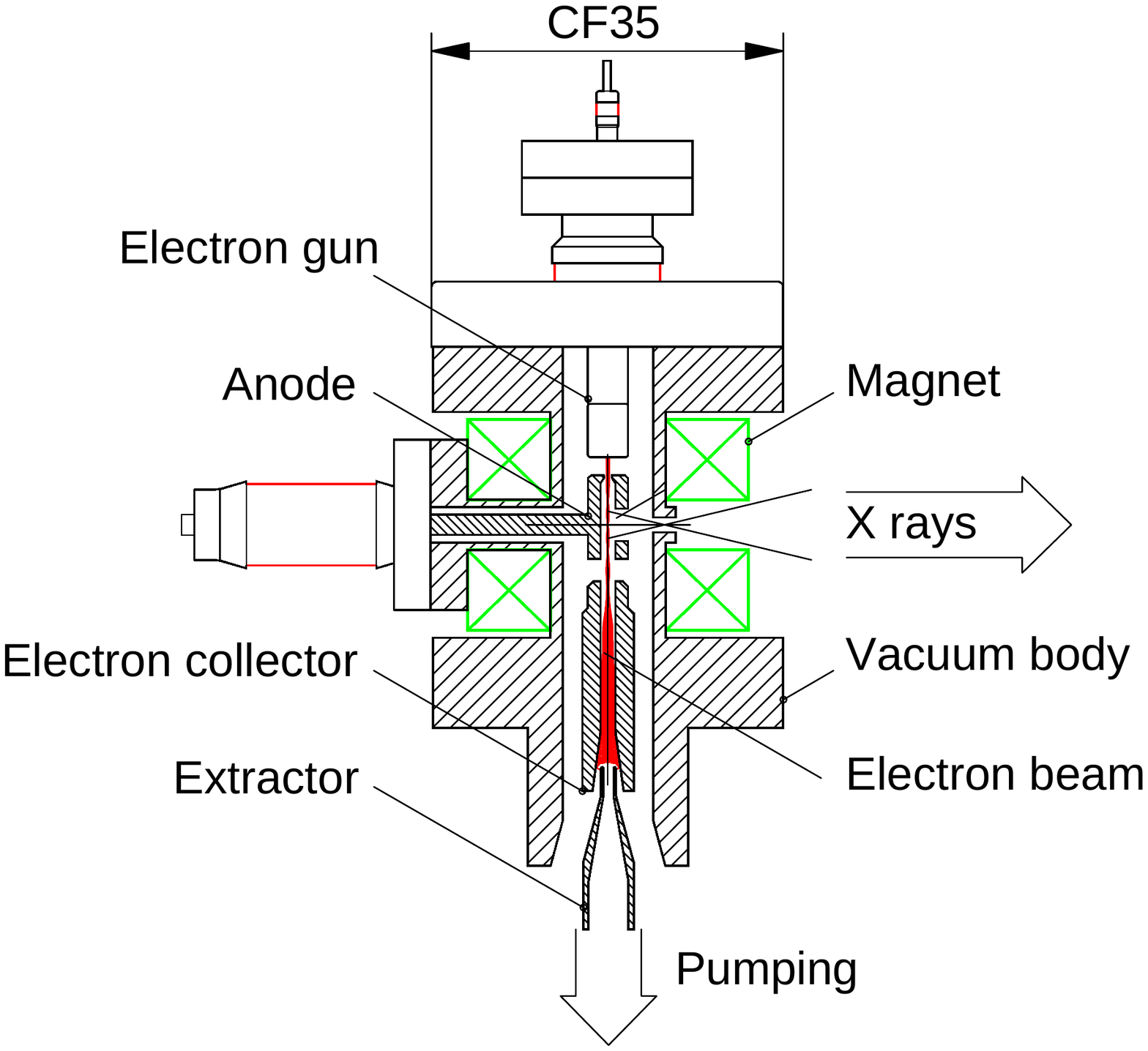}
\hfill
\includegraphics[width=0.45\columnwidth]{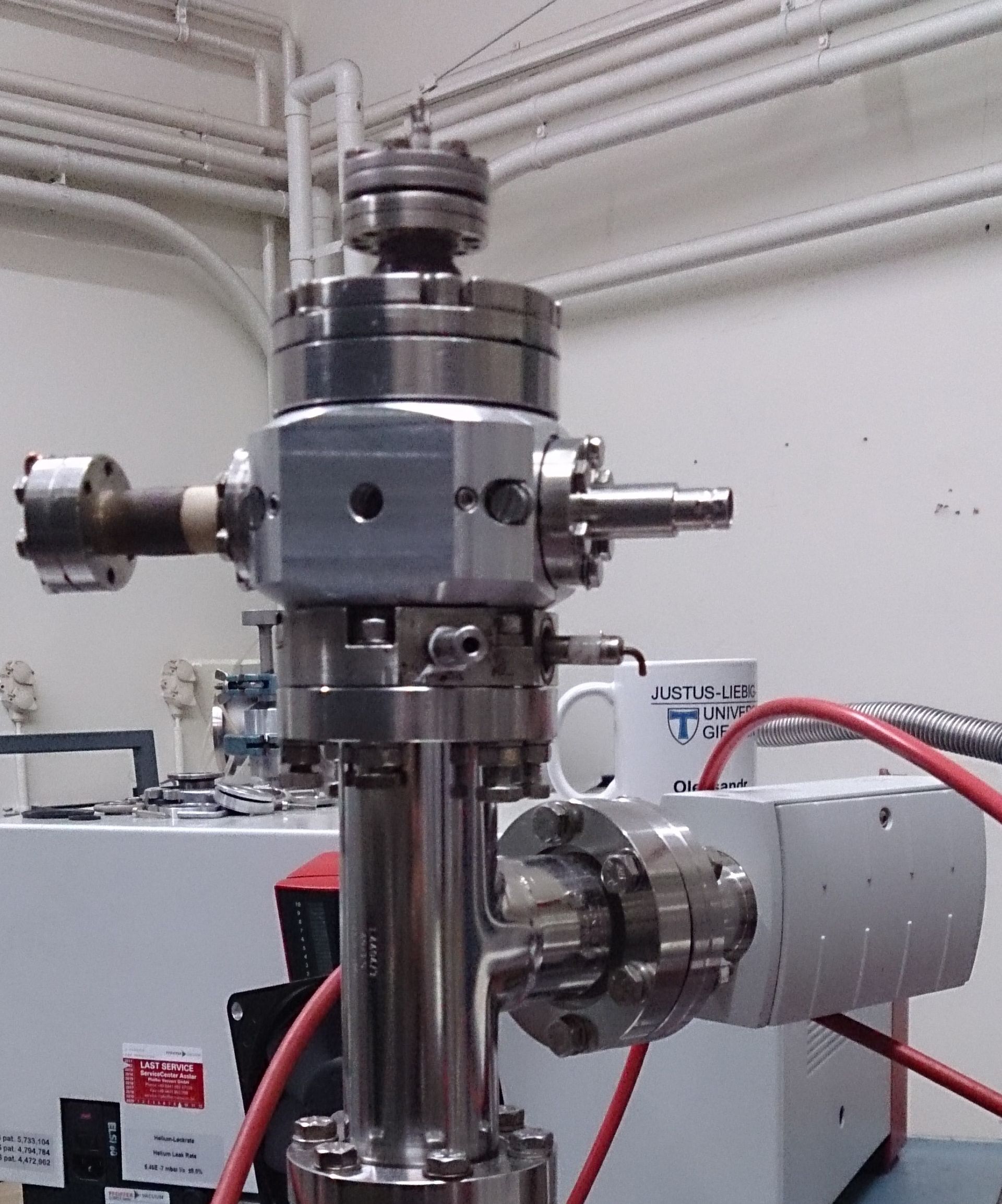}
\caption{\label{fig4} (Color online) MaMFIS-10: principle scheme of the construction and general view.}
\end{figure}

The principle sketch of construction and the general view for pilot model of the MaMFIS-10 are shown in Fig.~\ref{fig4}. Vacuum vessel of the ion source is equipped by standard 35-mm-ConFlat flanges, since the device does not have its own pumping system. Accordingly, the ion source can be installed at any vacuum system, depending on the purpose of application. Pumping is carried out along the $z$ axis.

The major part of the ion source is the magnetic focusing system, which provides the magnetic field of high strength over the short range. The magnetic field of solenoid type is created by permanent magnets with magnetization in the radial direction and consists of two parts. The length of gap between these parts in the $z$ direction is only 10 mm. The magnet system is located outside the vacuum vessel and does not have the iron pieces of complicated form inside of the vacuum volume. Two beryllium windows of 3-mm diameter are pasted directly on the vacuum body in the middle plane between the magnet parts. The electron beam formed by the electron gun is going through the anode to the electron collector. The extractor electrode stops the electrons and focuses the outgoing ions.

\section{Basic properties of MaMFIS-10}

The main objective of the experiment was to investigate efficiency of the local trap for the production of ions in high-charge states. The electron optics was tested for electron beam with the current of up to 50 mA within the energy range from 5 to 10 keV. The best results were achieved for the electron current of about 45 mA and the beam energy of 8.5 keV. In this case, the intercepted current is about 50 $\mu$A only. Therefore, further measurements were performed at these particular parameters of the electron beam. The device was running in the trapping mode without ion extraction.

The ionization processes in the local ion trap were studied experimentally by the method of x-ray spectroscopy. The theoretical analysis of x-ray radiation from the ion trap provides evidences of the production of highly charged ions. In order to determine the highest charge state achieved in the ion trap, we shall restrict ourselves to consideration of the high-energy band of radiative recombination only. The basic pro\-per\-ties of the ion source are tested with the mixture of iridium and cerium as the working substance, which is evaporated from the cathode.

\begin{figure}[bp]
\centering
\includegraphics[width=0.7\columnwidth]{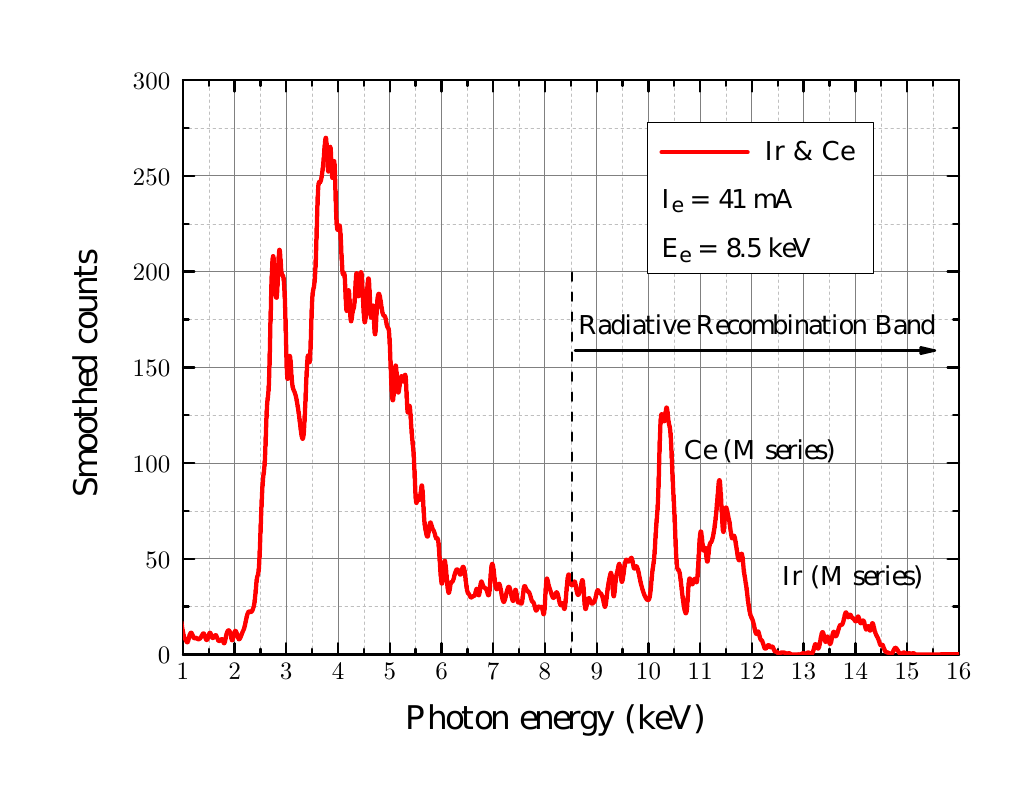}
\caption{\label{fig5} (Color online) X-ray spectrum of iridium and cerium ions (cathode materials) measured with the use of Si(Li) detector in one hour run.}
\end{figure}

\begin{figure}[hbtp]
\centering
\includegraphics[width=0.7\columnwidth]{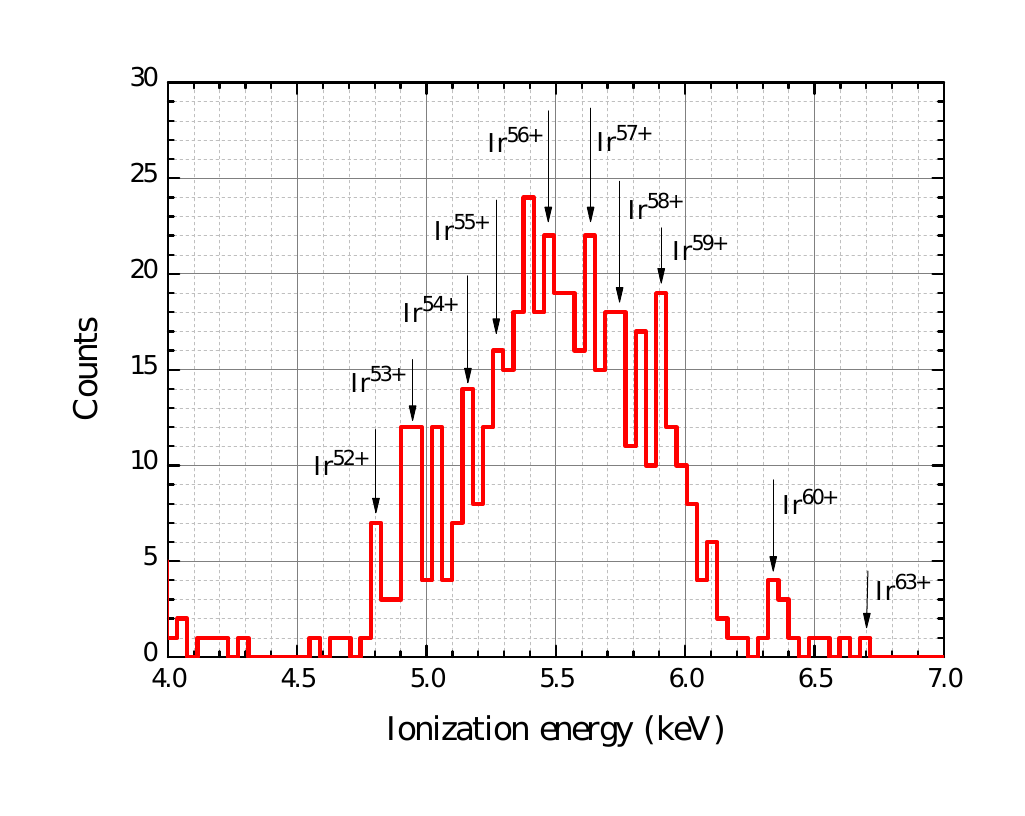}
\caption{\label{fig6} (Color online) X-ray emission due to radiative recombination into the M shell of iridium ions. The measurements are performed for the electron current $I_e$ of 41 mA, the energy $E_e$ of 8.5 keV, and the permanent pressure of the background gas (mainly H${}_2$) of $2\times 10^{-8}$ mbar.}
\end{figure}

In Fig.~\ref{fig5}, the peak near 14 keV corresponds to the M-series radiation due to recombination of iridium ions with the beam electrons. The analysis of charge states in this energy region is shown in Fig.~\ref{fig6}. Here and in the following, the arrows indicate the theoretical positions of radiative recombination peaks of multicharged ions with respect to the ionization energy. The ionization energy $E_i$ is defined as the difference between the energy of emitted x rays and the electron energy $E_e=8.5$ keV. The radiation of iridium ions with the charges of up to $q=+59$ can be reliably identified. Although the accurate analysis of this spectrum is known in the literature \cite{3}, here the spectrum is shown just to demonstrate the ionization ability of the ion source. The ionization factor realized in the MaMFIS is comparable to that obtained in the first EBIT \cite{4}.

\begin{figure}[tp]
\includegraphics[width=0.7\columnwidth]{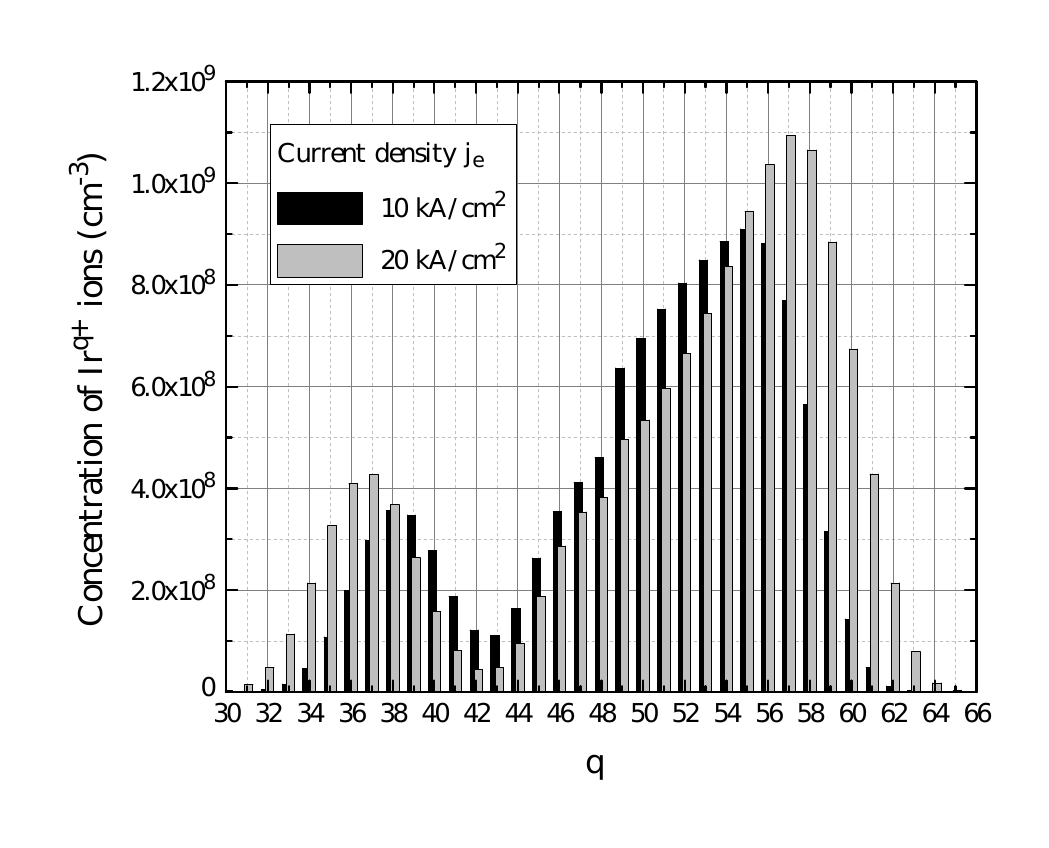}
\caption{\label{fig7} Theoretical predictions for the charge-state distributions of Ir ions. The electron beam energy is $E_e =8.5$ keV.}
\end{figure}

For determination of the electron current density, we have simulated the physical processes in the ion trap and made a comparison of the theoretical and experimental charge-state distributions. The computer simulations were performed by using the codes described in work~\cite{5} with the trap parameters corresponding to conditions of the experiment for two values of the current density $j_e$, namely, 10 and 20 kA/cm$^2$ (see Fig.~\ref{fig7}). The vacuum parameters were also chosen to match the experimental conditions. The permanent partial pressures are $5 \times 10^{-10}$ and $2 \times 10^{-8}$ mbar for neutral iridium and molecular hydrogen, respectively. A comparison of the theoretical and experimental charge-state distributions allows one to estimate the effective current density achieved in the ion source. It turns out to be larger than 10 kA/cm$^2$.

\section{Injection of working gas into the ion trap}

In the preliminary experiment with injection of working gas into the ion trap we are interested in the equilibrium microplasma. Due to ``evaporative cooling" in the local ion trap, confinement of ions with the charge lower than the charge of cathode materials becomes problematic \cite{4,6}.

When all transition processes of ionization, deionization, heating, and cooling are completed, the equilibrium plasma is established in local ion trap. In this case, the confinement time is not defined. Even in the basic conditions without injection of working gas, the ions produced from the residual substance can be captured into the local ion trap. The residual substance supplies the ions from the rest molecules in vacuum. In addition, the ions are also produced from the cathode materials due evaporation and sputtering processes. Usually, the cathode contains heavy elements, such as barium, tungsten, iridium and others. The ions of these materials being captured into the ion trap with extremely high electron current density increase their charges very rapidly. The amount of highly charged ions in the ion trap exceeds the amount of ions of lighter elements in lower charge states, because heavy highly charged ions of cathode materials push the ions of light working substance in relatively low charge states out of the trap. As a result, in the stationary plasma with steady-state temperature the amount of ions of light elements can become negligible.

The mechanism of ``evaporative cooling" for materials with large atomic numbers and permanent concentration of neutrals is very complicated, because it involves huge amount of ion-ion collision processes in the local ion trap. The computer simulation of these processes with use of the existing computer code \cite{5} is extremely time-consuming.

\begin{figure}[tbhp]
\includegraphics[width=0.7\columnwidth]{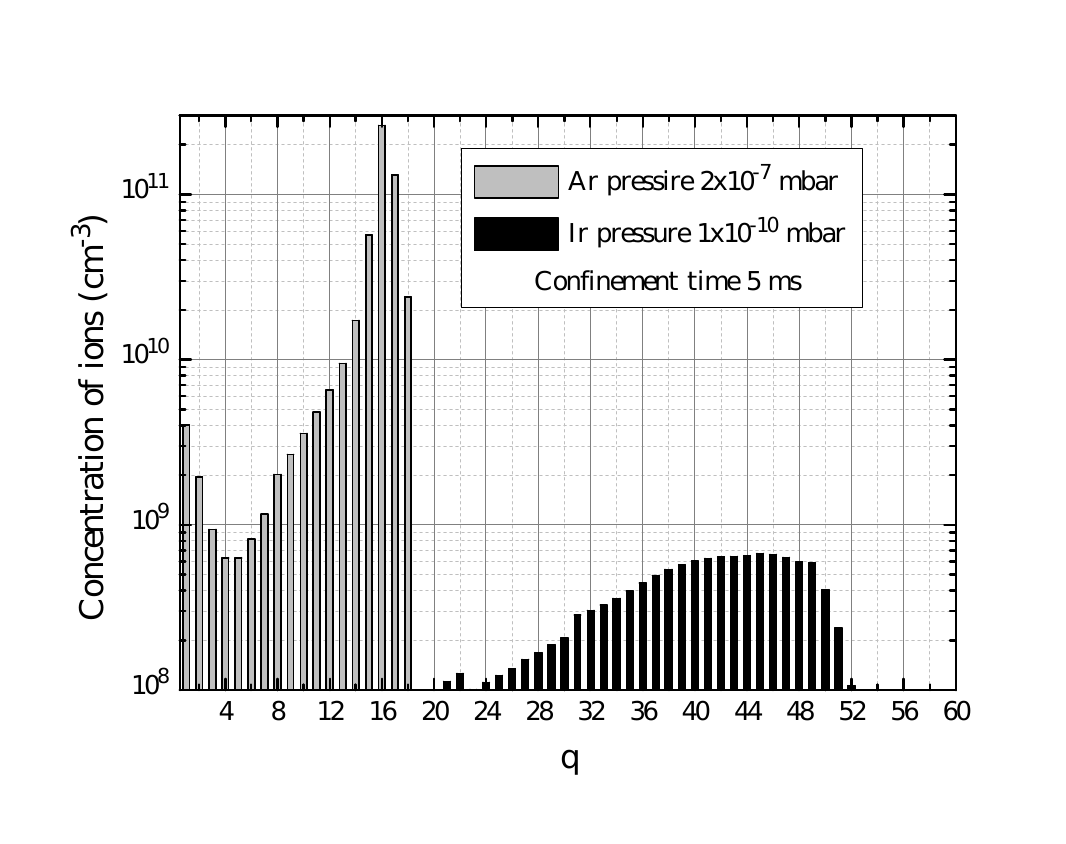}
\caption{\label{fig8} Charge-state distributions for ions of Ir and Ar after 5 ms of confinement.}
\end{figure}

\begin{figure}[tbhp]
\includegraphics[width=0.7\columnwidth]{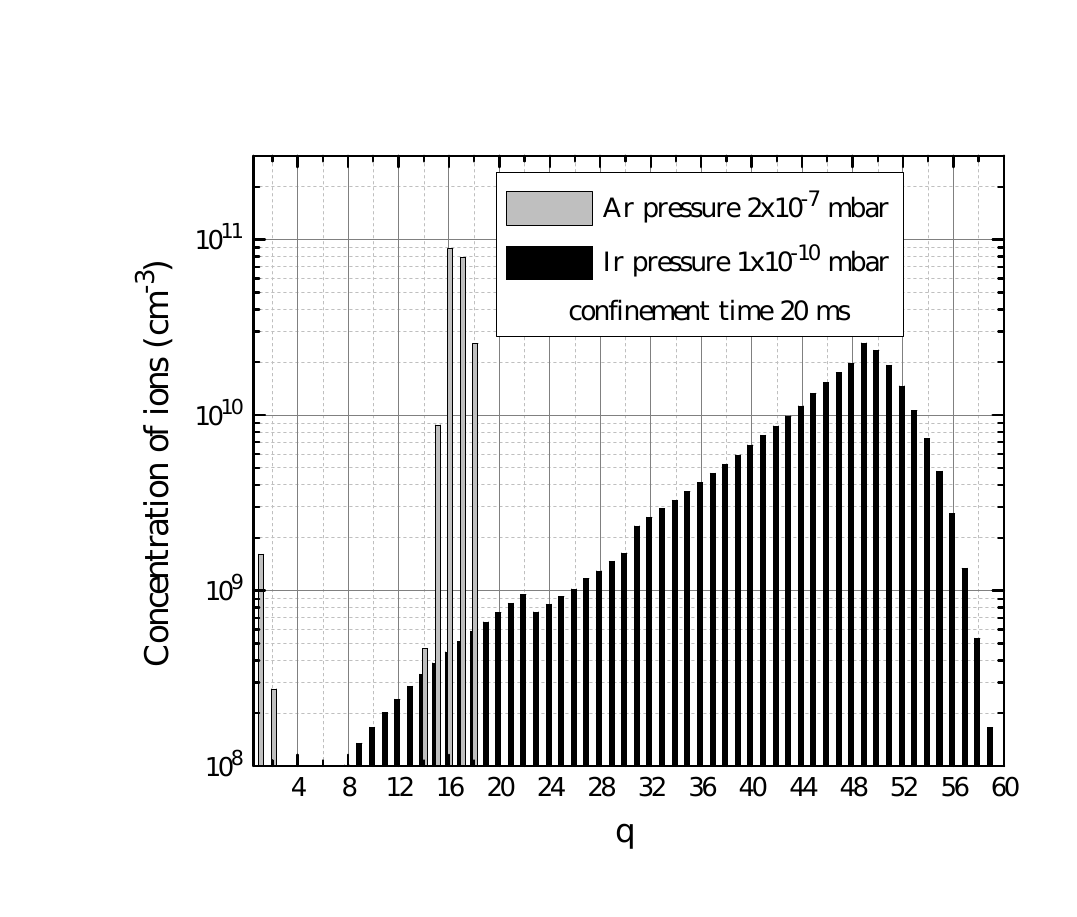}
\caption{\label{fig9} Similar to Fig.~\ref{fig8} for the confinement time of 20 ms.}
\end{figure}

Here we present theoretical computation of the ionization processes in the ion trap for two substances, Ir and Ar. In Figs.~\ref{fig8} and \ref{fig9}, charge-state distributions for ions of Ir-Ar mixture with the confinement time of 5 ms and 20 ms, respectively, are presented. The parameters of the local ion trap are the following: electron current is 50 mA, electron energy is 8.5 keV, electron current density is 20 kA/cm$^2$, and the trap length is 1 mm. Although the concentration of the neutral Ir atoms in 2000 times less than that of Ar atoms, even in such a short time of ionization the concentration of Ar ions decreases two times, while the concentration of Ir ions increases by a factor of thirty. In the equilibrium microplasma, the amount of Ar ions becomes almost zero.


There are two ways for detecting the x rays emitted by ions of light elements from the equilibrium plasma. The first way is to improve vacuum in the region of the electron gun. This tends to decrease the concentration of heavy ions in the trap due to reduction of the cathode sputtering. The second way is to increase the partial pressure of light elements in the region of the ion trap. This reduces the charges of heavy ions due the charge-exchange processes and, accordingly, increases the chance to observe the rest of light highly charged ions of working substance.

\begin{figure}[tbhp]
\centering
\includegraphics[width=0.7\columnwidth]{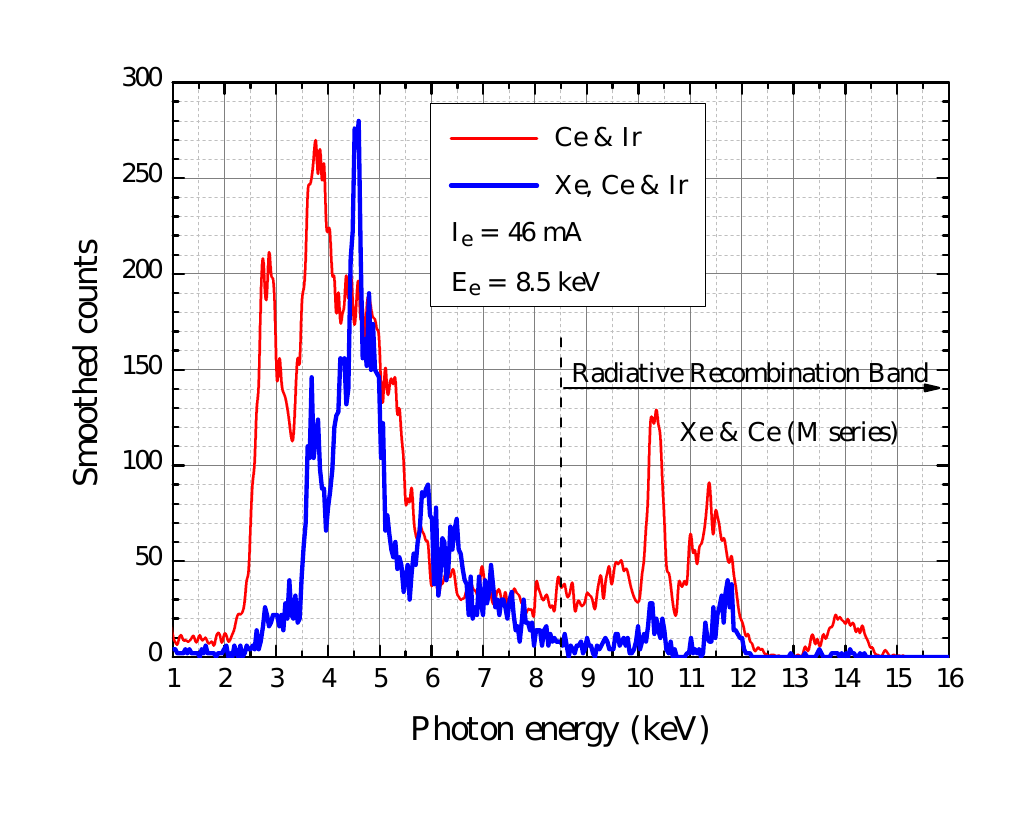}
\caption{\label{fig10} (Color online) X-ray spectrum of xenon, cerium, and iridium ions observed in the MaMFIS-10. Basic spectrum of the cathode materials is also given for comparison.}
\end{figure}

\begin{figure}[tp]
\centering
\includegraphics[width=0.7\columnwidth]{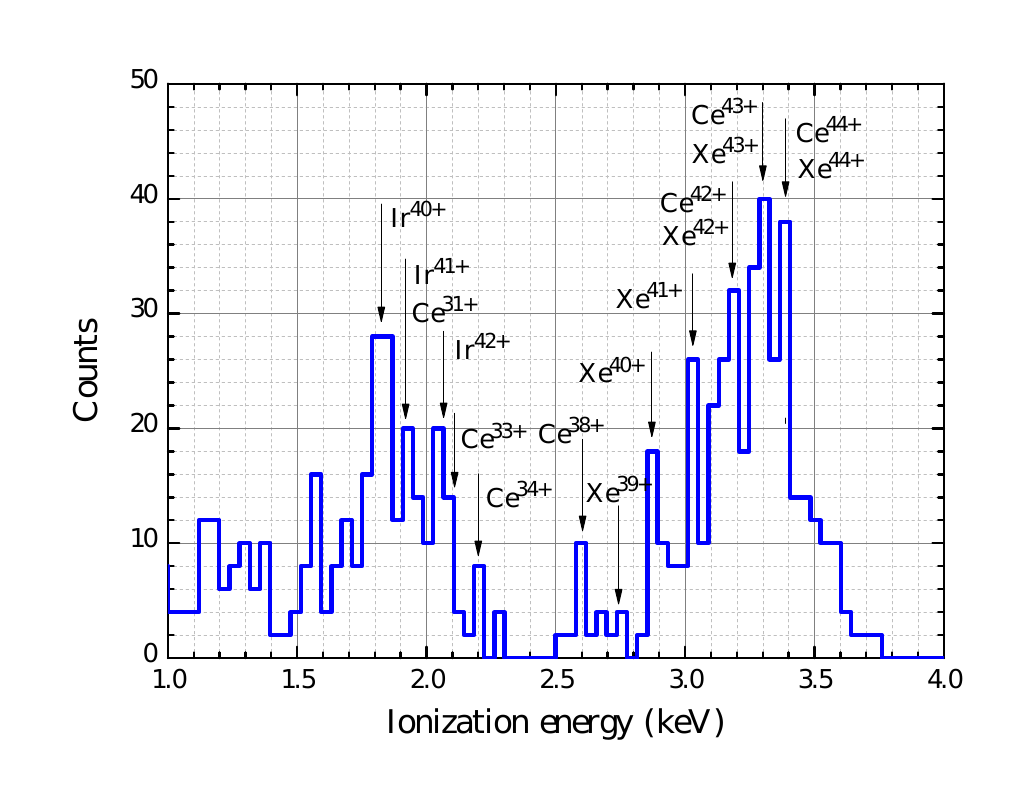}
\caption{\label{fig11} (Color online) X-ray emission of xenon, cerium, and iridium ions caused by the radiative recombination. The measurements are performed for the current $I_e$ of 46 mA and the current density $j_e$ of 8.5 keV. The total pressure in the ion trap is of about $2.2\times 10^{-8}$ mbar.}
\end{figure}

The result of injection of Xe ions can be seen in Fig.~\ref{fig10}, where the corresponding spectrum is given in   comparison with the basic spectrum emitted by the ions of Ir and Ce. Due to increase in the concentration of atoms in the trap region, the intensity of x-ray emission from highly charged Ir ions becomes relatively small. Identification of charge states of Xe ions by using the radiative recombination peaks is somewhat problematic, because the ionization M-shell energies for Xe and Ce ions are close to each other. Nevertheless, the difference of the total spectra indicates that some amount of Xe ions remains in the trap. The analysis of high-energy part of the spectrum allows one to identify the xenon ions with the charges of up to $q=+44$ (see Fig.~\ref{fig11}). The basic vacuum is estimated to be on the level of $10^{-8}$ mbar. This value is very approximate, because the direct measurement of vacuum in the local ion trap is impossible. The injection of xenon gas increases the pressure in the installation up to $2.2\times 10^{-8}$ mbar. Although highly charged ions of the cathode materials (Ir and Ce) are confined in the ion trap, the most of peaks in Fig.~\ref{fig11} are identified with emission of Xe ions.

\begin{figure}[tp]
\centering
\includegraphics[width=0.7\columnwidth]{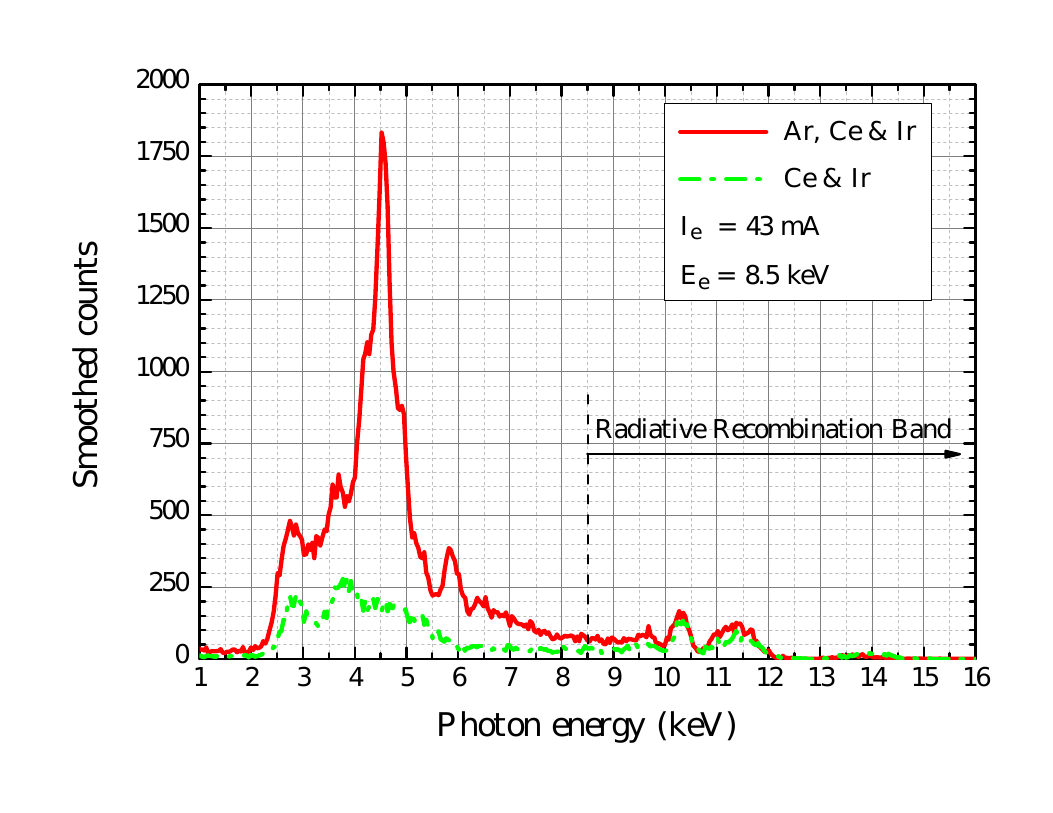}
\caption{\label{fig12} (Color online) X-ray spectrum of argon ions observed in the MaMFIS-10. The basic pressure (mainly due to molecular hydrogen) is $2.2 \times 10^{-8}$ mbar. The partial pressure of neutral argon is $1.2 \times 10^{-7}$ mbar.}
\end{figure}

\begin{figure}[bthp]
\centering
\includegraphics[width=0.7\columnwidth]{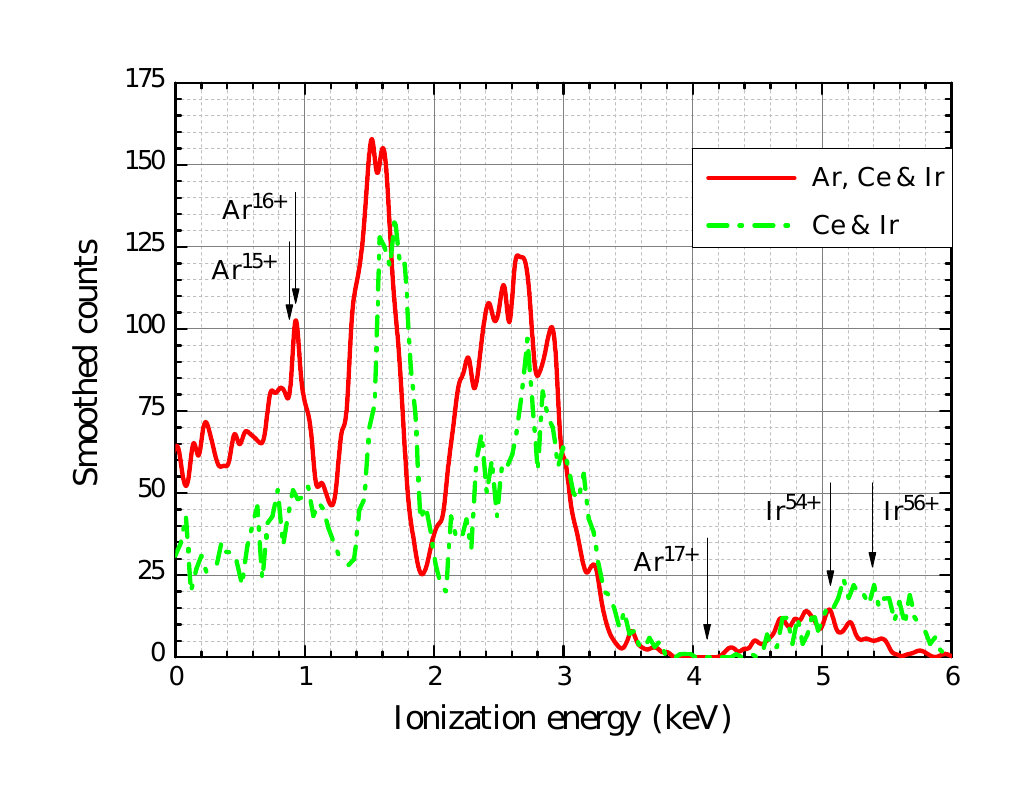}
\caption{\label{fig13} (Color online) X-ray emission of argon ions induced by the radiative recombination. The electron beam is characterized by the current $I_e$ of 43 mA and the current density $j_e$ of 8.5 keV.}
\end{figure}

In the case of injection of argon, the pressure of working gas was significantly increased up to the level of $1.2 \times 10^{-7}$ mbar. It enhanced the cathode sputtering and, subsequently, the penetration of heavy ions into the trap. The x-ray spectrum for mixture of the working gas and cathode materials is presented in Fig.~\ref{fig12}. The emission of cathode materials in the basic vacuum of $2.2\times 10^{-8}$ mbar is also measured for comparison. The x-ray spectrum due to the radiative recombination is analyzed in Fig.~\ref{fig13}. Despite a significant amount of heavy ions of iridium and cerium, the peak originated from the Ar$^{16+}$ ions is observed clearly. Note also that the emission spectrum of highly charged ions near $E_i$ of 5 keV is changed with increasing pressure of the working gas. Since the influence of charge-exchange processes is enhanced, the maximum of the charge-state distribution is shifted to the range of lower charge states. In particular, the maximum in photon counts of iridium ions is displaced from Ir$^{56+}$ to Ir$^{54+}$.

The most effective method for the production of highly charged ions of light elements can be realized in the running mode with control over the confinement time. According to theoretical predictions, the confinement time for the production of these ions is very short. For example, in electron beam with the current density $j_e$ of 10 kA/cm$^2$, the Ar$^{16+}$ ions can be prepared for the confinement time of less than 1 ms. During this period of time the ions of heavy elements do not have chance to replace the ions of lighter elements in the ion trap. Therefore, we conclude that the pulse running mode with a very short confinement time is necessary for the MaMFIS in the case of injection of light elements. For this purpose, the control of the local ion trap should be realized.

\begin{figure}[bthp]
\centering
\includegraphics[width=0.7\columnwidth]{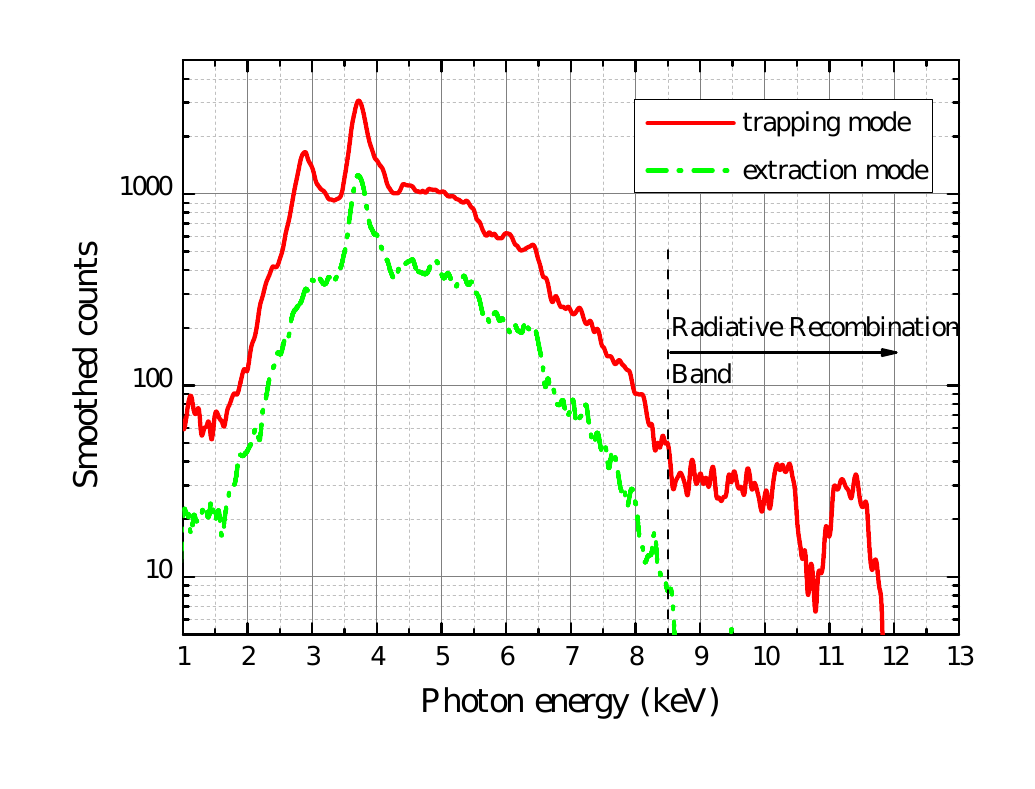}
\caption{\label{fig14} (Color online) X-ray spectra measured in trapping and extraction modes.}
\end{figure}

\section{Control of the local ion trap}

The control over depth of the local ion trap can be performed by changing the shape of the electron beam (see Fig.~\ref{fig1}). If the rippled electron beam is transformed into the smooth flow, the local potential well disappears. The smooth potential distribution allows for the axial extraction of ions. The x-ray spectrum of cathode materials was measured for both trapping and extraction running modes. In Fig.~\ref{fig14}, the solid curve corresponds to the trapping potential distribution (solid curve in Fig.~\ref{fig3}). The highly charged ions emit photons due to the radiative recombination, while the local ion trap exists. In this case, the potential of cathode is equal to the potential of focusing electrode. The decrease of the Wehnelt's potential by about 150 V causes transformation of the rippled electron beam into the smooth beam. Accordingly, the x-ray emission due to the radiative recombination is significantly suppressed (see dashed curve in Fig.~\ref{fig14}), since the highly charged ions are being extracted.

\section{Conclusions}

The method for the production of highly charged ions in local ion trap created by the rippled electron beam in a focusing magnetic field is realized. The highly charged ions such as Ir$^{59+}$ and Xe$^{44+}$ are produced in the room-temperature hand-sized device. The electron current density in the local ion trap is estimated to be of the order of 10 kA/cm$^2$. The experiments confirm this theoretical estimate. The novel ion source shows wide perspectives for laboratory investigations of the atomic and plasma physics.

\section*{ACKNOWLEDGMENTS}

The authors express deep gratitude to A.~M\"{u}ller for giving opportunity to test the MaMFIS in the laboratory of Institute of Atomic and Molecular Physics (Justus-Liebig University of Giessen), to O.~K.~Kultashev for his contribution to creation of the electronic optics, to A.~Borovik,~Jr. and K.~Huber for their support in x-ray measurements, to I.~V.~Kalagin for his contribution to development of computer codes, and to A.~Nukin for production of some details of the MaMFIS.

\end{document}